\documentclass{article}

\usepackage[final]{rml2018}

\usepackage[utf8]{inputenc} 
\usepackage[T1]{fontenc}    
\usepackage{hyperref}       
\usepackage{url}            
\usepackage{booktabs}       
\usepackage{amsfonts}       
\usepackage{nicefrac}       
\usepackage{microtype}      
\usepackage[pdftex]{graphicx}

\title{Enabling End-To-End Machine Learning Replicability: A Case Study in Educational Data Mining}

\author{
  Josh Gardner \\
  School of Information\\
  The University of Michigan\\
  \texttt{jpgard@umich.edu} \\
   \And
   Yuming Yang \\
  Department of Statistics\\
  The University of Michigan\\
  \texttt{yangym@umich.edu} \\
  \And
   Ryan S. Baker \\
  Graduate School of Education\\
  The University of Pennsylvania\\
  \texttt{rybaker@upenn.edu} \\
  \And
   Christopher Brooks \\
  School of Information\\
  The University of Michigan\\
  \texttt{brooksch@umich.edu} \\
}

\begin{document}

\maketitle

\begin{abstract}
The use of machine learning techniques has expanded in education research, driven by the rich data from digital learning environments and institutional data warehouses. However, replication of machine learned models in the domain of the learning sciences is particularly challenging due to a confluence of experimental, methodological, and data barriers. We discuss the challenges of end-to-end machine learning replication in this context, and present an open-source software toolkit, the MOOC Replication Framework (MORF), to address them. We demonstrate the use of MORF by conducting a replication at scale, and provide a complete executable container, with unique DOIs documenting the configurations of each individual trial, for replication or future extension at \url{https://github.com/educational-technology-collective/fy2015-replication}. This work demonstrates an approach to end-to-end machine learning replication which is relevant to any domain with large, complex or multi-format, privacy-protected data with a consistent schema.
\end{abstract}

\section{Introduction}

The repeated verification of scientific findings is central to the construction of robust scientific knowledge, particularly in a fast-growing field such as machine learning. This can take the form of (a) reproduction (reproducibility), using the original methods applied to the original data to reproduce the original results, and (b) replication (replicability), applying the original methods to \textit{new} data to assess the robustness and generalizability of the original findings. Since reproducibility is a necessary condition for replicability (an experimental procedure cannot be applied to \textit{new} data if the procedure cannot even be reproduced), any replication procedure also requires solving the problem of reproducibility.

In this work, we discuss the reproducibility crisis in machine learning, noting specific challenges faced by applied researchers in the learning sciences, particularly in the sub-fields of educational data mining and learning analytics. We argue that existing frameworks for reproducible machine learning, such as open code-sharing platforms and public code notebooks (such as those solicited by this workshop) are valuable steps, but are insufficient to fully address the challenges both within our subfield of interest and the broader machine learning community. In particular, we argue that code-sharing does not address the breadth of challenges -- experimental, methodological, and data -- we face as practitioners, as Section \ref{sec:barriers} details. 

Instead, we propose a paradigm of \textit{end-to-end} reproducibility for machine learning: fully reproducing (or replicating) the pipeline from raw data to model evaluation. End-to-end reproducibility is possible with current freely-available computing technologies, namely containerization, which can encapsulate an experiment as well as operating system and software dependencies in a single package. We describe a novel open-source platform for conducting reproducible end-to-end machine learning experiments on large-scale educational data, the MOOC Replication Framework (MORF), and describe additional benefits, beyond reproducibility, afforded by this platform in Section \ref{sec:advantages}.

\section{Prior Work}

\subsection{The Reproducibility Crisis in Machine Learning}~\label{sec:reproducibility-in-ml}
Much has been written about the reproducibility crisis in science, particularly in fields which conduct human subjects research such as social psychology. Recent empirical evidence has shown that issues with reproducibility are also widespread in the field of machine learning. In some ways, this realizes concerns that have been voiced in the field for nearly twenty years concerning the complex technical steps and dependencies involved in conducting a machine learning experiment which are rarely described or shared in adequate detail with existing publication methods. A recent survey of 400 research papers from leading artificial intelligence venues shows that none of the works surveyed document all aspects necessary to fully reproduce the work; only 20-30\% of the factors evaluated were adequately reported in the works surveyed \cite{Gundersen2017-zl}. In a case study of replication of deep reinforcement learning algorithms, \cite{Henderson2018-qd} show that the variance inherent to statistical algorithms, the use of different hyperparameter settings, and even different random number generation seeds contribute to a lack of reproducibility in machine learning research and have a direct impact on whether experimental results and baseline model implementations replicate. Olorisade et al. \cite{Olorisade2017-wv} survey 30 machine learning studies in text mining, and identify poor reproducibility due to lack of access to data, software environment, randomization control, and implementation methods. None of the 30 works surveyed provided source code, and only one of 16 applicable studies provided an executable, to reproduce their experiments.

These reproducibility issues are partly attributable to culture and convention. A survey of authors published in the Journal of Machine Learning Research found that roughly one third intentionally did not make their implementations available, for reasons including a lack of professional incentives, a reluctance to publish messy code, and the convention that doing so is optional \cite{Sonnenburg2007-da}. Peng \cite{Peng2011-az} observes that only five of 125 published articles in the journal \textit{Biostatistics} have passed the (voluntary) reproducibility review since its inception two years prior, yet considers this effort ``successful'' compared to the reproducibility of previous work.

As big data and machine learning permeate disciplines, this crisis in replication has also affected other fields of study, such as the learning sciences. This is especially relevant in cases of very large datasets where the majority of learning is computer-mediated, such as in Massive Open Online Courses (MOOCs). For example, \cite{Gardner2018-wo} showed that a large-scale replication of machine learning models led to substantially different conclusions about the optimal modeling techniques, with several findings replicating significantly in the \textit{opposite} direction of the original study (which was conducted on only a single MOOC). Khajah et al. \cite{Khajah2016-md} attempted to replicate the findings of the ``deep knowledge tracing'' method introduced in \cite{Piech2015-yy}, and demonstrated that much simpler baseline methods could achieve equivalent performance, and that the initial performance gains demonstrated in the original work were at least partially due to data leakage.

\begin{figure}
    \centering
    \includegraphics[width = \textwidth]{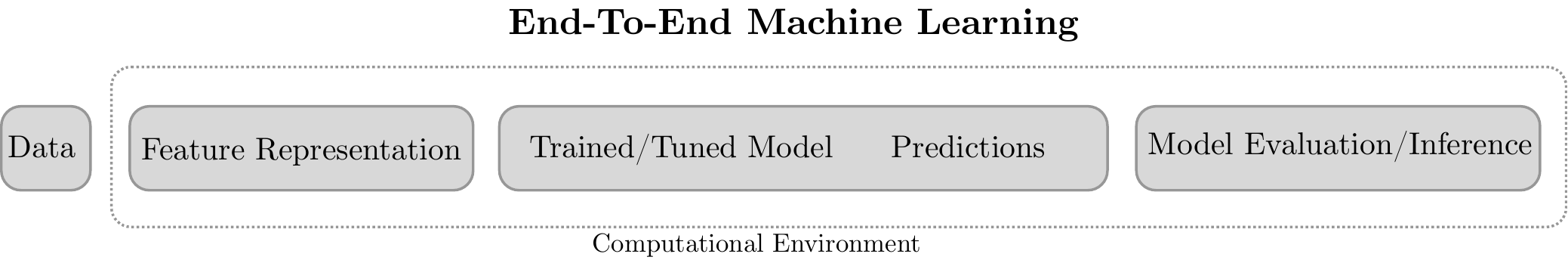}
    \caption{End-to-end reproducibility requires addressing data, technical, and methodological issues with reproducibility. Replication of the computational environment, in particular, is key to replicating this complete pipeline from raw data to results.}
    \label{fig:endtoend}
\end{figure}

\subsection{Existing Tools for Reproducible Machine Learning}~\label{sec:existing-tools}
An exhaustive survey of tools and platforms to support reproducible machine learning research is beyond the scope of this work. However, we include a survey of those tools which most closely align with our interests in building machine learned models for predictive analytics in education.

OpenML is ``an open, organized, online ecosystem for machine learning'' and allows users to create data science pipelines to address specific ``tasks'', such as classification and clustering \cite{Van_Rijn2013-af}. OpenML requires users to run their analyses locally and upload the results. OpenML's workflow may contribute to ``openness,'' but does not ensure reproducibility, as it relies on user-submitted code to fully reproduce the results which are generated and submitted by users. OpenML functions as a code-, data-, and results- sharing platform, but these analyses are subject to key limitations (dependency management, code rot) described in more detail in Section \ref{sec:barriers}. Finally, OpenML presumes data can be shared freely for replication, an often incorrect presumption in the field of the learning sciences, where institutional and governmental regulations and policies manage how data can be shared.

The OpenAI Gym is an open-source interface for developing and comparing reinforcement learning algorithms \cite{Brockman2016-sx}. Its wide use for both teaching and research serve as an example of how a field can create and adopt shared tools to support researchers' needs and support reproducibility. For example, \cite{Henderson2018-qd} uses this platform for its reproducible baseline models. However, OpenAI Gym supports only reinforcement learning (not other machine learning tasks such as supervised learning).

Several publishing platforms dedicated to reproducible computational research also exist, such as ReScience \footnote{ \url{http://rescience.github.io/about/} }, CodaLab \footnote{ \url{http://codalab.org/}}, and WholeTail \cite{Brinckman2018-gq}. These platforms unify code, data, computation, and presentation in a single location. CodaLab and WholeTail also use Docker containerization to ensure reproducibility. However, these platforms generally provide no support or restricted-access data; for example, WholeTail specifically limits its scope to public (non-restricted) datasets \cite{Brinckman2018-gq}.

Each of these platforms is an important step toward reproducible machine learning research, and many of them address key barriers. However, these tools are insufficient for many types of machine learning tasks, including supervised learning with large-scale behavioral data from MOOCs. In particular, none of these platforms supports replication where the underlying data sets are privacy-restricted and cannot be publicly shared. In educational data, many of the types of unanonymizable data that are privacy-restricted are also necessary for analysis (such as the text of discussion forum postings, IP addresses, or student names). Such restrictions are also likely to drive away machine learning researchers from working with this data, as gaining access to unprocessed raw educational data can be difficult or impossible without close collaborators and strong institutional support. Even with institutional support, government regulations such as the Family Educational Rights and Privacy Act (FERPA) may restrict or complicate data sharing.

\section{Barriers to End-to-End Replication in Machine Learning}~\label{sec:barriers}
The replication crisis is the result of a confluence of forces which must be collectively addressed in order to achieve end-to-end reproducibility. We group these challenges into experimental, methodological, and data challenges. No existing solution discussed in Section \ref{sec:existing-tools} currently addresses all three barriers.

Experimental challenges with reproducibility relate to reproducing the exact experimental protocol.\footnote{The term ``experimental reproducibility'' is adopted from \cite{Gundersen2017-zl}.} Many have advocated for the open sharing of code as a potential solution to address technical issues with reproducibility \citep[e.g.][]{Stodden2013-bh, Peng2011-az}. However, while necessary, code-sharing is insufficient to guarantee reproducibility in machine learning research. For example, Collberg et al. \cite{Collberg2014-oh} showed that the published code accompanying  20\% of their large sample of 613 published computer systems papers failed to build or run, and in total, it was not possible to verify or reproduce 75.1\% of studies surveyed using the artifacts provided in publication.

Even when code is available, other technical issues can prevent reproducibility in computational research workflows \cite{Donoho2015-aq, Kitzes2017-pf}. Buckheit and Donoho \cite{Buckheit1995-zw} noted over 20 years ago that the complete software environment is a necessary condition for reproducing computational results; however, the open sharing of such environments along with the published results of scientific work remains rare. These include \textit{code rot}, in which code becomes non-functional or its functionality changes as the underlying dependencies change over time (for example, an update to a data processing library which breaks backwards compatibility, or a modified implementation of an algorithm which changes experimental results), as well as \textit{dependency hell}, in which configuring the software dependencies necessary to install or run code prevents successful execution \cite{Boettiger2015-cu}. This complex web of interdependencies is rarely described or documented in published machine learning and computational science work \cite{Donoho2015-aq, Gundersen2017-zl, Olorisade2017-wv}. 

Fortunately, modern containerization tools were developed to resolve such technical issues in software development contexts \cite{Merkel2014-xu}. Docker containers, for instance, are frequently used in both industrial software applications as well as computational and computer systems research \cite{Boettiger2015-cu, Cito2016-ej, Jacobsen2015-bq}. A major advantage of Docker over simple code-sharing is that Docker containers fully reproduce the entire execution environment of the experiment, including code, software dependencies, and operating system libraries. These containers are much more lightweight than a full virtual machine, but achieve the same level of reproducibility \cite{Merkel2014-xu, Jacobsen2015-bq}. While some of the platforms discussed in Section \ref{sec:existing-tools} utilize containerization, these platforms ``hide'' this functionality from the user, assembling containers from user-submitted code. This severely limits users' ability to fully leverage containerization by building the complex, customized environments many machine learning experiments may require.

Methodological challenges to reproducibility reflect the methods of the study, such as its procedure for model tuning or statistical evaluation. Existing work on reproducibility largely focuses on strictly technical challenges, but (as our experiment in Section \ref{sec:morf} shows), methodological issues are at least as important. Methodological challenges include the use of biased model evaluation procedures \cite{Cawley2010-la, Varma2006-by}, the use of improperly-calibrated statistical tests for classifier comparison \cite{Dietterich1998-vh}, or ``large-scale hypothesis testing'' where thousands of hypotheses or models are tested at once, despite the fact that most multiple testing corrections are not appropriate for such tasks \cite{Efron2016-to}. A machine learning version of these errors is seen in massive unreported searches of the hyperparameter space, and in ``random seed hacking'' wherein the random number generator itself is systematically searched in order to make a target model's performance appear best or a baseline model worse \cite{Henderson2018-qd}. Replication platforms can address methodological issues -- working toward what \cite{Gundersen2017-zl} terms \textit{inferential reproducibility}-- by architecting platforms which support effective methodologies and adopt them by default, effectively nudging researchers to make sound choices. 

Data reproducibility concerns the availability of data itself. In many domains, making raw data available is more an issue of convention than a true barrier to reproducibility. However, in the case of educational data mining, data are governed by strict privacy regulations which protect the privacy of student education records. Similar restrictions affect many other fields, from the health sciences to computational nuclear physics \cite{Kitzes2017-pf}. As a result, researchers are often legally prohibited from making their data available. Efforts such as the Pittsburgh Science of Learning Center DataShop \cite{Koedinger2010-yg} and the HarvardX MOOC data sets \cite{HarvardX2014-ds} have attempted to address this problem in education by only releasing non-identifiable data, but many analyses require the original, unprocessed data for a full replication. Indeed, restricted data sharing is one of the main factors (in our experience) hindering generalizability analysis in educational data mining: investigators are generally limited to one or two courses worth of data (e.g. the courses they instruct), and models are often overfit to these datasets.

\section{The MOOC Replication Framework}~\label{sec:morf}
MOOC Replication Framework (MORF) is a Python toolkit, accompanied by a platform-as-a-service (the ``MORF Platform''), which collectively address the challenges faced by researchers studying large-scale online learning data noted above.\footnote{The MORF website, which includes documentation and short tutorials, is at \url{https://educational-technology-collective.github.io/morf/}} MORF addresses (i) experimental barriers to replication via the use of containerization; (ii) methodological issues by providing support for true holdout sets and reliable model evaluation; (iii) data issues by providing a large-scale dataset while preventing the download of sensitive information (currently, this includes the complete raw data exports from over 270 MOOCs offered by the University of Michigan and the University of Pennsylvania). Additionally, MORF eases the computational expense of conducting such research at scale by providing nearly an order of magnitude greater computational infrastructure than any of the platforms discussed in Section \ref{sec:existing-tools}, and out-of-the-box parallelization to utilize it.

Users submit jobs to the MORF Platform using a simple Python API. A short, four-line ``controller'' script is used to guide the execution of each stage of the end-to-end supervised learning pipeline (extract, train, test, and evaluate). The use of controller scripts is a best practice in computational research \cite{Kitzes2017-pf}, and MORF's combination of containerization and controller scripts allow the user to manage low-level experimental details (operating system and software dependencies, feature engineering methods, and statistical modeling) by constructing the Docker container which is submitted to MORF for execution, while MORF manages high-level implementation details (parallelization, data wrangling, caching of results). The containers used to execute each job run on MORF are persisted in MORF's public Docker Cloud repository, and the configuration file and controller scripts are persisted in Zenodo and assigned a unique Digital Object Identifier (DOI). This yields a reproducible end-to-end pipeline that is highly flexible, easy to use, and computationally efficient.

\subsection{Experimental Reproducibility via Containerization}
MORF's uses containerization to resolve many of the experimental challenges described in Section \ref{sec:barriers}. The Docker containers submitted to MORF fully encapsulate the code, software dependencies, and execution environment required of an end-to-end machine learning experiment in a single file, ensuring end-to-end reproducibility and enabling sharing of the containerized experiment.  Building Docker containers requires only a single Dockerfile (akin to a makefile) which contains instructions for building the environment. This imposes minimal additional burden on researchers relative to configuring, programming, and executing an experiment, but achieves a considerable increase in reproducibility. While other existing machine learning research platforms sometimes utilize docker ``under the hood'' for reproducibility, this limits users' ability to configure or share these environments. We are not aware of any which allow users to build and submit Docker images directly for execution.

As part of MORF, we are assembling an open-source library of pre-built Docker containers to replicate experiments conducted on MORF to serve as shared baseline implementations. These containers can be loaded with a single line of code, allowing the research community to replicate, fork, interrogate, modify, and extend the results presented here\footnote{The experiment presented below can be loaded by running \texttt{docker pull themorf/morf-public:fy2015-replication} in the terminal of any computer with Docker installed.}.

\subsection{Methodological Reproducibility via Platform Architecture}
MORF provides sensible default procedures for many machine learning tasks, such as model evaluation. For example, MORF avoids the use of cross-validation for model evaluation: The prediction tasks to which most MOOC models aspire are prediction of \textit{future} student performance (i.e., in an ongoing course where the true labels -- such as whether a student will drop out -- are unknown at the time of prediction). As such, using cross-validation within a MOOC session, when the outcome of interest is accuracy on a \textit{future} MOOC session, provides an unrealistic and potentially misleading estimate of model performance. Prior work has demonstrated that cross-validation provides biased estimates of independent generalization performance \cite{Varma2006-by}, and in the MOOC domain, that cross-validation can produce biased estimates of classification performance on a future (unseen) course \cite{Whitehill2017-tt}. Adopting more effective model evaluation techniques by default requires no additional work for MORF users, and ensures that work produced on the MORF platform follows effective model evaluation procedures.

\subsection{Data Reproducibility via Execute-Against Access}

MORF achieves data reproducibility while also meeting data privacy restrictions by providing strictly ``execute-against'' access to underlying data. Most MOOCs are generated by a small number of platforms (e.g. Coursera, edX), and all courses from a given platform use publicly-documented data schemas \citep[e.g.][]{Coursera2013-jh}. Thus, users can develop experiments using their own data from a given platform -- or even the public documentation -- and then submit these experiments for MORF to execute against \textit{any} other course from that platform. This enables MORF to currently provide an interface to over 200 unique sessions of more than 70 unique courses offered by two different institutions on the Coursera platform, and to execute containerized experiments against this data in a secure, sandboxed environment, by utilizing the shared, public schema of MOOC datasets. These shared public data schemas also ensure that existing experiments in MORF can be replicated against new data (from the same MOOC platform) as it becomes available.

\section{Case Study in Replication with MORF}
We demonstrate the use of MORF through a case study. We replicate a comparison conducted in \cite{Fei2015-ea}, which compares several machine learning algorithms using a set of seven activity features (e.g. number of lecture videos viewed, quizzes attempted, and discussion forum posts for each student) over each week in a MOOC in order to predict a binary dropout label indicating whether a user showed activity in the final week of a course.

This study is an ideal candidate for replication because it compares several models (effectively testing many pairwise hypotheses) using cross-validation on only a single dataset. This testing of many hypotheses/comparisons, with only a single observation for each, can lead to poor methodological reproducibility and provides no information about the variability of the estimates, relative to their magnitude. Particularly because this experiment was concerned with empirical performance (in order to inform future ``early warning'' dropout prediction systems), obtaining an accurate estimate of models' expected performance on \textit{future} course sessions across a large, representative dataset can provide insight into the generalizability of these findings. This work is indicative of prediction tasks in the fields of learning analytics and educational data mining (see \cite{Gardner2018-lp} for a more thorough discussion). 

\begin{figure}
    \centering
    \includegraphics[width = \textwidth]{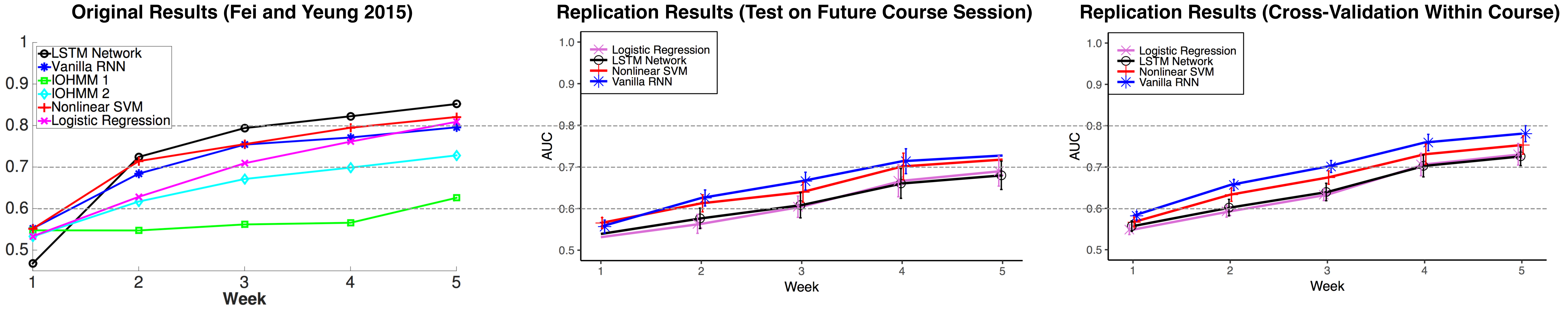}
    \caption{Original results from \cite{Fei2015-ea} (left) and replication results using the MOOC Replication Framework evaluated using a held-out future course session (center) and cross-validation (right). 95\% confidence intervals shown. IOHMM models not replicated due to lack of an open-source implementation which supported prediction.}
    \label{fig:replication_results}
\end{figure}

We replicated the original experiment across 45 unique MOOCs using MORF. Results are shown in Figure \ref{fig:replication_results}.\footnote{We would like to acknowledge the helpful assistance of the original authors in conducting this replication.} 

The original work argued that a Long Short-Term Memory (LSTM) neural network model `` beats the ... other proposed methods by a large margin'' \citep[][pp. 1]{Fei2015-ea}. Our results show, however, that (1) LSTM is actually one of the worst-performing models, with the lowest average performance of any model tested in weeks 4 and 5; (2) in most cases, the 95\% confidence intervals for algorithm performance overlap (indicating no significant difference); and (3) observed performance of all models is lower than observed in \cite{Fei2015-ea}, particularly in later weeks of the course.

\begin{figure}
    \centering
    \includegraphics[width = \textwidth]{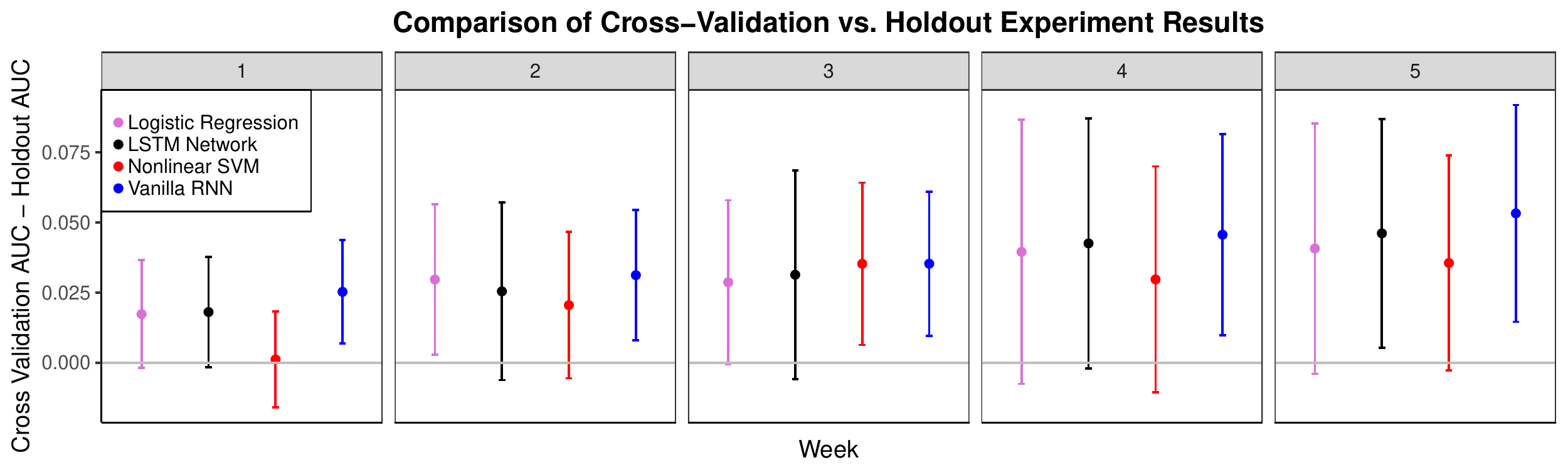}
    \caption{Comparison of AUC estimates from identical experiments using a holdout vs. cross-validation evaluation architecture. These results show a persistent positive bias when cross-validation is used to estimate predictive performance on a future, held-out session of a MOOC.}
    \label{fig:cv_holdout_comparison}
\end{figure}

We hypothesize that (1) may be due to overfitting on the original training data. Particularly when using cross-validation on a single dataset with a highly flexible model such as LSTM, the original work was quite susceptible to overfitting. Overfitting seems particularly likely because no procedure for selecting hyperparameters was reported in the original work, and some relevant hyperparameter settings for the final LSTM model (e.g. batch size) were not reported at all. These hyperparameters were not available even after correspondence with the authors, who did not record them and no longer had the original code available (which itself points to the need for reliable long-term reproducibility solutions such as MORF). Point (2) shows the advantage of using MORF's large data repository, which allows us to observe variability in each algorithms' performance across many MOOCs. This result suggests that while differences in average performance may exist, these are too small to be interpreted as genuine and not spurious -- particularly in light of the results shown in Figure \ref{fig:cv_holdout_comparison}, which shows that the differences due to cross-validation bias are, in many cases, larger than the observed differences between algorithms. Finally, (3) is likely due to the cross-validation used in the original experiment (as opposed to the holdout architecture used in MORF). The original experiment allowed models to train and test on subsamples from the same overarching population, which the holdout architecture used in MORF requires the (more challenging) task of predicting student performance on a future course session -- which is the true target of prediction in this work.

The right panel of Figure \ref{fig:replication_results} demonstrates the results of an identical replication experiment, but evaluated using five-fold cross-validation (as in \cite{Fei2015-ea}) instead of the future held-out course session used in the center panel. The contrast between the center (holdout) and right (cross-validation) panels demonstrates the optimistic bias which can be introduced by evaluating generalization performance via cross-validation without the use of an independent hold-out set. This comports with previous results demonstrating that the bias of performance estimates when models are optimized over cross-validation folds can often exceed the difference between learning algorithms \cite{Cawley2010-la}. These results are further demonstrated by Figure \ref{fig:cv_holdout_comparison}, which shows a small but persistent positive bias for model evaluation performed by cross-validation versus the ``true'' performance on a future course session. A two-sided Wilcoxon signed rank test of a null hypothesis of equivalence between the holdout and cross-validated experimental results rejected with $p < 2.2 \times 10^{-16}$.

The results of this experiment demonstrate the importance of using large, diverse datasets for machine learning experiments, and of performing multiple experimental replications. Additionally, these results demonstrate that simpler models -- such as RNN, radial SVM, and logistic regression -- achieve equivalent or better performance relative to LSTM on our specific task of MOOC dropout prediction. Our results, which are contrary to the findings of the original study, also suggest that further replication is necessary to identify the most effective algorithms for MOOC dropout prediction, as we perform no hyperparameter tuning and only replicate the original models and features.

\section{Additional Advantages of Reproducible Machine Learning Frameworks}\label{sec:advantages}

Much prior work on reproducibility has focused on verification -- ensuring that published results are true and can be reproduced. However, end-to-end reproducible machine learning frameworks, such as MORF, provide benefits beyond mere verification, including:

\textbf{``Gold standard'' benchmarking:} open replication platforms allow for the comparison of results which were previously not comparable, having been conducted on different data. The use of such benchmarking datasets has contributed to the rapid advance of fields such as computer vision (e.g. MNIST, IMAGENET), natural language processing (Penn Tree Bank, Brown corpus), and computational neuroscience (openFMRI). These datasets have been particularly impactful in fields where it is difficult or expensive to collect, label, or share data (as is the case with MOOC data, due to legal restrictions on sharing and access). These help to evaluate the ``state of the art'' by providing a common performance reference which is currently missing in many fields.

\textbf{Shared baseline implementations:} We noted above that variability in so-called ``baseline'' or reference implementations of prior work has contributed to concerns about reproducibility in the field \cite{Henderson2018-qd}. By providing fully-executable versions of existing experiments, MORF ameliorates these issues, allowing for all future work to properly compare to the exact previous implementation of a baseline method.

\textbf{Forkability:} containerization produces a ready-made executable which fully encompasses the code and execution environment of an experiment. These can be readily shared and "forked" across the scientific community, much in the same way code is "forked" from a git repository. This allows machine learning scientists to build off of others' work by modifying part or all of an end-to-end pipeline (for example, by experimenting with different statistical algorithms but using the same feature set as a previous experiment) within the same software ecosystem. 

\textbf{Generalizability analysis:} Each successive replication of an experiment provides information about its generalizability. Evaluating the generalizability of experiments has been a challenge in MOOC research to date, where studies conducted on small, restricted, and often homogenous datasets tend to dominate the literature. These challenges are shared by many other fields where data is scarce and sharing is restricted, such as nuclear physics or medical research \cite{Kitzes2017-pf}. When containerized end-to-end implementations are available, replicating these analyses on new data -- even data which are not publicly available but share the schema of the original data -- becomes as straightforward as running the containerized experiment against new data.

\textbf{Sensitivity Analysis:} This technique, used widely in Bayesian analysis, evaluates how changes to the underlying assumptions or hyperparameters affect model fit and performance. Such an evaluation can provide useful information about a model's robustness and potential to generalize to new data. Without being able to fully reproduce a model on the original data, sensitivity analyses are not possible. In MORF, such analyses can be conducted by simply forking and modifying the containerized version of the original experiment, then re-executing it against the same data. These analyses can also include so-called \textit{ablation analyses}, wherein individual components are removed from a model to observe their contribution to the results, as well as \textit{slicing analyses}, where fine-grained analysis of performance across different subgroups (e.g. demographic groups) is explored \cite{Sculley2018-md}.

\textbf{Full Pipeline Evaluation:} Each stage of an end-to-end machine learning experiment (feature extraction, algorithm selection, model training, model evaluation) can be done in many different ways. Each stage also affects the others (for example, some algorithms might perform best with large feature spaces; others might perform poorly with many correlated features). However, current research usually evaluates only one or two components of this pipeline (e.g. training several algorithms and tuning their hyperparameters on a fixed feature set). Not only are the remaining stages often described in poor detail or not at all \cite{Gundersen2017-zl}; such work also leaves future researchers unable to evaluate the synergy between different aspects of the end-to-end pipeline in a published experiment (for example, exploring whether an algorithm's performance improves with a different feature set). MORF fully encapsulates this end-to-end pipeline for a given experiment and it makes it available for modification to any other researcher.

\textbf{Meta-Analysis:} While meta-analyses are common in fields with robust empirical research bases, such analyses have been less common in the field of machine learning, which has an emphasis on novelty. The open availability of executable machine learning experiments affords detailed meta-analyses by providing complete results of all modeling stages for meta-analysis.

\section{Conclusion}
While MORF currently supports MOOC data, the broader workflow it uses to enable end-to-end machine learning reproducibility is domain-agnostic, and the problems it seeks to address are faced by many fields which use machine learning. We believe that the paradigm of end-to-end reproducibility can be adopted by any domain which uses large, complex, multi-format data. A platform such as MORF is fully capable of being adapted to such domains, particularly those which use restricted-access data. We hope that future work utilizes and extends our open-source implementation, and that MORF contributes to the construction of robust machine learning research.

\bibliographystyle{plain}
\bibliography{morf_icml_2018}







\end{document}